\documentclass[fleqn,12pt,twoside]{article}
\usepackage{espcrc1}
\usepackage{bm}
\usepackage{epsfig}
\usepackage{psfig}

\usepackage{graphicx}
\newcommand{\AmS}{{\protect\the\textfont2
  A\kern-.1667em\lower.5ex\hbox{M}\kern-.125emS}}

\title{Elliptic flow from partially thermalized heavy-ion 
       collisions\thanks{This work was supported by the U.S. 
                         Department of Energy
                         under Contract No. DE-FG02-01ER41190.}
}

\author{Ulrich Heinz and Stephen M.H. Wong\\[1ex]
        Physics Department,
        The Ohio State University, Columbus, OH 43210}
       
\begin{document}

% typeset front matter
\maketitle

\begin{abstract}\noindent
We study to what extent the measured elliptic flow at RHIC constrains 
viscous deviations from ideal hydrodynamics. We solve a toy model where 
only transverse momenta are thermalized while the system undergoes 
longitudinal free-streaming. We show that RHIC data exclude such 
a model and thus require fast {\em 3-dimensional} thermalization.
\end{abstract}

%%%%%%%%%%%%%%%%%%%%%%%%%%%%%%%%%%%%%%%%%%%%%%%%%%%%%%%%%%%%%%%%%%%%%%%%%%
\section{MOTIVATION}
\label{sec1}
%%%%%%%%%%%%%%%%%%%%%%%%%%%%%%%%%%%%%%%%%%%%%%%%%%%%%%%%%%%%%%%%%%%%%%%%%%

One of the most important discoveries at RHIC so far has been the
strong elliptic flow generated in non-central collisions 
\cite{v2STAR,v2PHENIX,v2PHOBOS}. For not too peripheral collisions
the measured elliptic flow coefficient $v_2(p_\perp)$ at midrapidity 
almost exhausts the hydrodynamic limit \cite{O92,ourv2,v2massdep,Teaneyv2}
up to transverse momenta $p_\perp\leq 2$\,GeV, and in this entire domain
the dependence of $v_2$ on the mass of the emitted particles 
\cite{v2STAR,PHENIX_qm02} accurately follows the hydrodynamically predicted 
pattern \cite{v2massdep}. This means that the distribution of the momenta 
of well over 99\% of the emitted particles is accurately described by (ideal) 
hydrodynamics.

Why is this so important? The initial transverse momentum distribution of 
the particles generated by the colliding nuclei is locally 
isotropic. Only their {\em spatial} distribution in the transverse plane 
is initially deformed (for $b{\,\ne\,}0$). Interactions among the 
produced quanta are required to transfer this spatial anisotropy to 
momentum space. A non-vanishing $v_2$ is thus an unambiguous signature
for reinteractions in the produced matter, and the observed large $v_2$ 
values prove that the initial parton liberation process is separated from 
the experimentally observed final state by a violently interacting stage 
of dynamical evolution. Consequently, there is {\em a priori} little 
reason to expect that calculable properties of the early matter formed 
in the reaction zone immediately after nuclear impact (such as, for 
example, the central rapidity density and the shapes of the rapidity and 
transverse momentum distributions of the produced gluons \cite{Kharzeev}) 
have any direct relationship with the corresponding experimentally 
observed values. The evolution of spectral shapes due to the strong 
rescattering must be taken into account, and even though adiabatic 
cooling and the build-up of collective flow move the spectral slopes
in opposite directions it is highly unlikely that these effects cancel 
completely. Elliptic flow itself is an example for a {\em qualitative} 
change of the spectra between particle formation and decoupling.

Microscopic studies \cite{ZGK99,Molnar} show that $v_2(p_\perp)$ 
is a monotonic function of the mean free path, 
$\lambda{\,\sim\,}(\sigma\rho)^{-1}$; the hydrodynamic limit is 
approached {\em from below} as $\lambda{\,\to\,}0$. The observation that 
at RHIC $v_2(p_\perp)$ almost exhausts the hydrodynamic limit thus
appears to force the conclusion \cite{v2STAR,ourv2,v2massdep,Teaneyv2,HK02}
that the fireballs formed in Au+Au collisions at RHIC thermalize fast
and efficiently. Since the creation of elliptic flow is driven by the 
spatial deformation of the reaction zone which quickly decreases
either spontaneously via free-streaming \cite{ourv2} or (more
rapidly) as a result of the flow anisotropy itself \cite{Sorge}, $v_2$ 
is sensitive to the very early collision stage. Within hydrodynamics 
the RHIC $v_2$ data can only be reproduced by assuming thermalization 
at times $\tau_{\rm th}{\,\leq\,}1$\,fm/$c$ (we use 
$\tau_{\rm th}{\,=\,}0.6$\,fm/$c$ \cite{ourv2,v2massdep,HK02}). As the 
hydrodynamic limit of $v_2$ can only be reached from below and the data 
almost exhaust it, we can use the hydrodynamic model to estimate the 
energy density at thermalization. One finds values \cite{ourv2} at least 
an order of magnitude above the critical one for hadronization, 
$\epsilon{\,\simeq\,}1$\,GeV/fm$^3$. The thermalized state formed 
${\sim\,}1$\,fm/$c$ after nuclear impact must therefore have been a 
{\em quark-gluon plasma} which, according to the hydrodynamical 
simulations, lives for several fm/$c$ before hadrons first appear. 
Unless the presented chain of arguments leading to the 
conclusion of early thermalization can be broken, the implication is 
unavoidable that at RHIC a well-developed quark-gluon plasma has been 
created.

The purpose of the work reported here \cite{HW02} is to
try to poke holes into the early thermalization argument. 
It was already pointed out by Ollitrault \cite{O92} that $v_2$ is
sensitive to the stiffness of the equation of state (EOS) of the
thermalized matter and increases monotonically with the sound velocity
$c_s^2{\,=\,}\partial P/\partial e$. Might it be possible to trade 
off thermalization against a stiffer EOS? The concrete idea 
studied by us \cite{HW02} was that perhaps the thermalization 
of the (on average much larger) longitudinal momenta of the liberated 
partons takes longer than transverse thermalization, resulting in a 
smaller longitudinal than transverse thermal pressure. As shown by 
Teaney at this conference, this is similar to the expected effects 
from shear viscosity on the longitudinal hydrodynamic expansion 
\cite{Teaney}. If we exaggerate a bit and idealize the model by 
assuming collisionless free-streaming with boost-invariant initial 
conditions in the longitudinal direction coupled with complete local 
thermalization of the transverse momenta, we can write down an analytic 
expression for the phase-space distribution function in terms of 
macroscopic parameters for which we can analytically derive macroscopic 
equations of motion which generalize the usual ideal hydrodynamic 
equations. In this idealization there is no longitudinal pressure at 
all; all the hydrodynamic work goes in the transverse direction. Can 
we generate larger elliptic flow in this way? If yes, the RHIC data 
would no longer saturate the theoretical limit and the fast 
thermalization argument would break down. We'll see that it 
doesn't.

%%%%%%%%%%%%%%%%%%%%%%%%%%%%%%%%%%%%%%%%%%%%%%%%%%%%%%%%%%%%%%%%%%%%%%%%%%
\section{A TRANSVERSALLY THERMALIZED MODEL (TTHM)}
\label{sec2}
%%%%%%%%%%%%%%%%%%%%%%%%%%%%%%%%%%%%%%%%%%%%%%%%%%%%%%%%%%%%%%%%%%%%%%%%%%
%\subsection{Distribution function}
%\label{sec2a}
%%%%%%%%%%%%%%%%%%%%%%%%%%%%%%%%%%%%%%%%%%%%%%%%%%%%%%%%%%%%%%%%%%%%%%%%%%

With boost-invariant initial conditions, assuming that all particles
originate at $t{\,=\,}0$ from $z{\,=\,}0$ (i.e. infinitely Lorentz 
contracted colliding nuclei), an appropriate ansatz for the phase-space
distribution of massless gluons is
 \begin{equation}
 \label{eq1}
   f({\bm{x}},{\bm{k}},t)=\frac{\tau_0}{\tau} \,
            \delta(y-\eta)\,\frac{1}{\gamma_\perp}\,
            \frac{1}{e^{k{\cdot}u/T}-1}\,.
 \end{equation}
Here $\tau{\,=\,}\sqrt{t^2{-}z^2}$, $\eta{\,=\,}\frac{1}{2}\ln\frac{t{+}z}
 {t{-}z}$, and $y{\,=\,}\frac{1}{2}\ln\frac{k^0{+}k_z}{k^0{-}k_z}$.
We use longitudinal boost-invariance to parametrize the local flow as 
$u^\mu=\gamma_\perp(\cosh\eta,\bm{v}_\perp,\sinh\eta)$. The transverse 
flow velocity $\bm{v}_\perp$ with $\gamma_\perp{\,=\,}1/\sqrt{1{-}v_\perp^2}$
and the ``transverse temperature'' $T$ are functions of the longitudinal 
proper time $\tau$ and the transverse position $\bm{x}_\perp$, but 
independent of space-time rapidity $\eta$. Due to the free-streaming 
constraint $\delta(y{-}\eta)$ the exponent of the Bose distribution 
reduces to $k{\cdot}u{\,=\,}\gamma_\perp k_\perp 
 (1-\hat{\bm{k}}_\perp{\cdot}\bm{v}_\perp)$, showing thermalization of 
only the transverse momenta.

%%%%%%%%%%%%%%%%%%%%%%%%%%%%%%%%%%%%%%%%%%%%%%%%%%%%%%%%%%%%%%%%%%%%%%%%%%
%\subsection{Macroscopic dynamics}
%\label{sec2b}
%%%%%%%%%%%%%%%%%%%%%%%%%%%%%%%%%%%%%%%%%%%%%%%%%%%%%%%%%%%%%%%%%%%%%%%%%%

Inserting this ansatz into the kinetic definition of the gluon energy 
momentum tensor, $T^{\mu\nu}(\bm{x},t) = \nu_g \int 
\frac{d^3k}{(2\pi)^3}\,\frac{k^\mu k^\nu}{k^0}\, 
f(\bm{x},\bm{k},t)$ ($\nu_g=16$ is the gluon spin-color degeneracy), 
one finds
 \begin{equation}
 \label{eq3}
    T^{\mu\nu}=\Bigl(e+P_\perp\Bigr)u^\mu u^\nu 
              - P_\perp \Bigl[g^{\mu\nu} + n^\mu n^\nu 
              + v_\perp(u^\mu m^\nu{+}m^\mu u^\nu)\Bigr]
 \end{equation}
with the {\em equation of state} (EOS)   
$e(\bm{x}_\perp,\tau) = 2\, P_\perp(\bm{x}_\perp,\tau)
   = \nu_g (\tau_0/\tau) (\pi^2 T^4(\bm{x}_\perp,\tau)/60)$.
The difference between longitudinal and transverse momenta requires
two additional vectors to decompose $T^{\mu\nu}$, 
$n^\mu{\,=\,}(\sinh\eta,0,0,\cosh\eta)$ and $m^\mu{\,=\,}\gamma_\perp
 (v_\perp\cosh\eta,\hat{\bm{v}}_\perp,v_\perp\sinh\eta)$ with 
$n{\cdot}m{\,=\,}n{\cdot}u{\,=\,}m{\cdot}u{\,=\,}0$. The resulting
extra terms relative to the ideal fluid decomposition can be viewed 
as viscous corrections \cite{Teaney}. The EOS is consistent with a 
vanishing trace of $T^{\mu\nu}$ for massless particles and the absence 
of longitudinal pressure; the corresponding sound velocity 
$c_s{\,=\,}1/\sqrt{2}$ is larger than for an ideal gluon gas, i.e. the 
EOS is stiffer.

With the decomposition (\ref{eq3}) the macroscopic equations of motion
$\partial_\mu T^{\mu\nu}{\,=\,}0$ become
 \begin{equation}
 \label{eq5}
   \frac{\partial T^{0\nu}}{\partial \tau} +
            \frac{T^{0\nu}}{\tau} + \nabla^j_\perp T^{j0\nu} = 0,
   \qquad \nu=0,1,2,3, \qquad j=1,2,3.
 \end{equation}
The only difference from the analogous ideal hydrodynamic equations 
\cite{ourv2} is the absence of a term $-P/\tau$ on the right hand side 
of the $\nu{\,=\,}0$ equation, reflecting work done by the longitudinal 
pressure. Standard hydrodynamic codes for the transverse expansion of 
systems with longitudinal boost invariance can thus be used to solve 
these equations. 

%%%%%%%%%%%%%%%%%%%%%%%%%%%%%%%%%%%%%%%%%%%%%%%%%%%%%%%%%%%%%%%%%%%%%%%%%%%%%%
\section{RESULTS}
\label{sec3}
%%%%%%%%%%%%%%%%%%%%%%%%%%%%%%%%%%%%%%%%%%%%%%%%%%%%%%%%%%%%%%%%%%%%%%%%%%%%%%

Lacking longitudinal pressure, the energy density in TTHM decreases more 
slowly with $\tau$ than in ideal hydrodynamics (HDM). However, the larger 
transverse pressure performs more work in the transverse directions, 
resulting in stronger radial and elliptic flow. As a consequence, 
initial conditions which in HDM lead to a consistent description of the 
data produce in TTHM much too flat transverse momentum spectra. The 
reproduction of the central collision data via TTHM thus requires retuned 
initial conditions. This retuning is facilitated by the numerical 
observation \cite{HW02} that for massless 
particles the TTHM dynamics leads to completely time-independent 
transverse momentum spectra; transverse flow buildup thus {\em exactly} 
compensates for cooling! We have no analytical explanation for this fact. 
It implies that the {\em final} slope of the transverse momentum spectrum 
is given by the {\em initial} temperature; steeper final spectra thus 
require a reduced initial temperature. To maintain the measured 
normalization $dN/dy$ one then must also increase the initial 
longitudinal volume, by increasing the starting time $\tau_0$ for the
TTHM dynamics. This implies {\em late} transverse thermalization (and, 
of course, an even later longitudinal one).

To obtain the same final spectra and radial flow in TTHM as in HDM tuned 
to RHIC data, we find that we should start the transverse expansion about 
10 times later with a $\approx 15-20$ times lower initial energy density. 
The total remaining time until freeze-out is thereby shortened considerably, 
and even though we do not take into account the decrease of the spatial 
deformation of the reaction zone by transverse free-streaming prior to 
the onset of transverse TTHM dynamics, we find that now considerably less 
elliptic flow (only about half as much as in HDM) is generated (Fig.~1). 
As the data almost exhaust the HDM values, TTHM underpredicts them by 
about a factor two, both for the $p_\perp$-integrated elliptic flow 
$v_2$ (Fig.~1a) and for the $p_\perp$-slope of the differential elliptic 
flow $v_2(p_\perp)$ (Fig.~1b). The TTHM model is thus experimentally 
excluded; only a model such as HDM with a considerable degree of 
early longitudinal thermalization can describe the RHIC measurements. 

%%%%%%%%%%%%%%%%%%%%%%%%%%%%% Fig. 1 %%%%%%%%%%%%%%%%%%%%%%%%%%%%%%%%%%%%%%%%
\begin{figure}[htb]
\vspace*{-4mm}
\begin{minipage}[t]{79mm}
\psfig{bbllx= 29pt, bblly=44pt, bburx=316pt, bbury=265pt,
             file=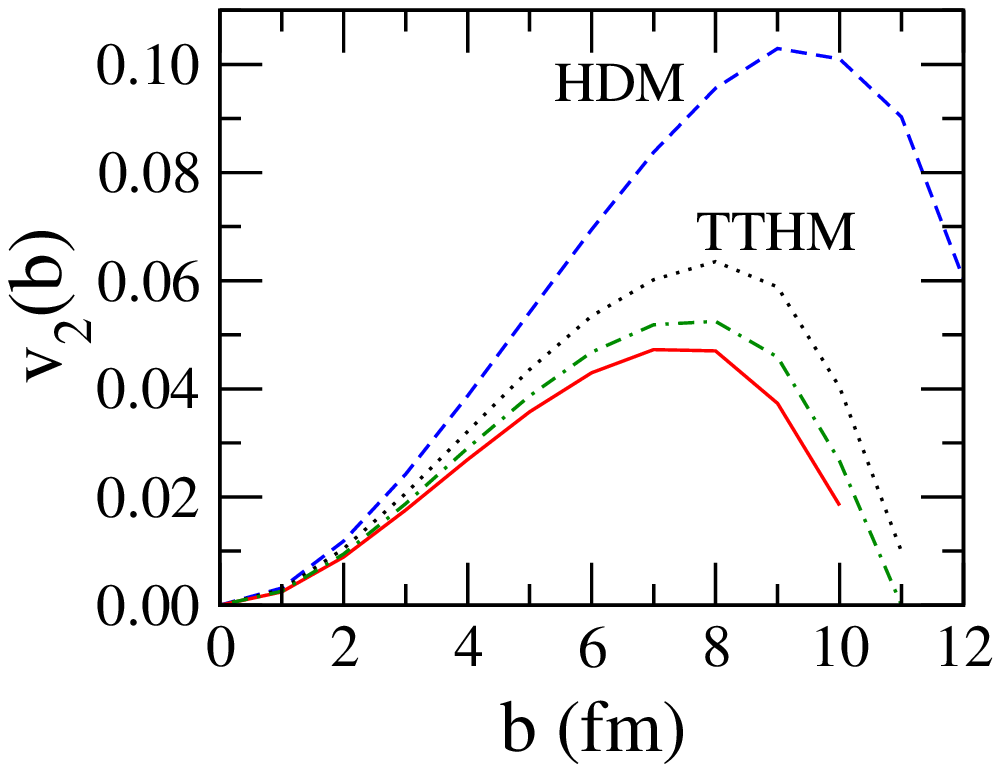, width= 7.9cm, height=6.5cm, clip=}
\label{fig:1a}
\end{minipage}
\hspace{\fill}
\begin{minipage}[t]{79mm}
\psfig{bbllx= 29pt, bblly=42pt, bburx=312pt, bbury=267pt,
             file=fig1b.ps, width= 7.9cm, height=6.5cm, clip=}
\label{fig:1b}
\end{minipage}
\vspace*{-1.4cm}
\caption{$v_2$ as a function of impact parameter $b$ (left) and of 
$p_\perp$ for two values of $b$ (right) from TTHM (solid) and HDM
(dashed). For details see Ref. \cite{HW02}.}
\end{figure}
\vspace*{-7mm}
%%%%%%%%%%%%%%%%%%%%%%%%%%%%%%%%%%%%%%%%%%%%%%%%%%%%%%%%%%%%%%%%%%%%%%%%%%%%%

In closing we note that for massless bosons $v_2(p_\perp{\,\to\,}0)$ 
approaches a non-zero value (Fig.~1b). This reflects the singular nature 
of the Bose distribution at $p_\perp{\,=\,}0$ \cite{HW02}.
We suggest to use this as a test for thermalization of small-$p_\perp$
gluons in parton cascades and of direct photons in experiment. 
   
%%%%%%%%%%%%%%%%%%%%%% References %%%%%%%%%%%%%%%%%%%%%%%%%%%%%%%%%%%%%%

\end{document}